\def\year{2019}\relax
\documentclass[letterpaper]{article} 
\usepackage{subcaption}
\usepackage{spconf}
\usepackage{times}  
\usepackage{helvet}  
\usepackage{courier}  
\usepackage{url}  
\usepackage{graphicx}  
\usepackage{multirow}
\usepackage{color}
\usepackage{cite}
\usepackage{amsmath}
\usepackage{amssymb}
\begin{document}
%
\title{Learning Disentangled Representations for Timber and Pitch\\ in Music Audio}
\name{Yun-Ning Hung$^1$, Yi-An Chen$^2$ and Yi-Hsuan Yang$^1$}
\address{$^{1}$ Research Center for IT Innovation, Academia Sinica, Taiwan \\
$^{2}$ KKBOX Inc., Taiwan\\
\tt \{biboamy,yang\}@citi.sinica.edu.tw, annchen@kkbox.com}

\maketitle
\begin{abstract}
Timbre and pitch are the two main perceptual properties of musical sounds. Depending on the target applications, we sometimes prefer to focus on one of them, while reducing the effect of the other. 
Researchers have managed to hand-craft such timbre-invariant or pitch-invariant features using domain knowledge and signal processing techniques, but it remains difficult to disentangle them in the resulting feature representations. 
Drawing upon state-of-the-art techniques in representation learning, we propose in this paper two deep convolutional neural network models for learning disentangled representation of musical timbre and pitch. 
Both models use encoders/decoders and adversarial training to learn music representations, but the second model additionally uses skip connections to deal with the pitch information. 
As music is an art of time,
the two models are supervised by frame-level instrument and pitch labels using a new dataset collected from MuseScore. 
We compare the result of the two disentangling models with a new evaluation protocol called ``timbre crossover,'' which leads to interesting applications in audio-domain music editing. 
Via various objective evaluations, we show that the second model can better change the instrumentation of a multi-instrument music piece without much affecting the pitch structure. 
By disentangling timbre and pitch, we envision that the model can contribute to generating more realistic music audio as well.
\end{abstract}

\section{Introduction}
\label{sec_intro}
Timbre and pitch are the two main perceptual properties of musical sounds. For a musical note, they refer to the perception of sound quality and frequency of the note, respectively. For a musical phrase, the perception of pitch informs us the notes and their ordering in the phrase (e.g., Do-Do-So-So-La-La-So), whereas the perception of timbre informs us the instruments that play each note. 
Timbre and pitch are interdependent, but they can also be \emph{disentangled}---we can use different instruments to play the same note sequence, and the same instrument to play different note sequences. 
While listening to music, human beings can selectively pay their attention to either the timbre or pitch aspect of music. 
For AI applications in music, we hope machines can do the same.

Preserving the characteristics of one property while reducing those of the other has been studied in the literature.
When the goal is to build a computational model that recognizes the note sequences (e.g., for tasks such as query by humming~\cite{ghias95mm} and cover song identification~\cite{serra08taslp}), we need a feature representation of music that is not sensitive to changes in timbre and instrumentation. In contrast, in building an instrument or singer classifier,
we may want to focus more on timbre rather than pitch.

The pursuit of such timbre- or pitch-invariant features has been mostly approached with domain knowledge and signal processing techniques \cite{mueller11jstsp,muellerEwert10taslp}. 
However, the effectiveness of these features is usually evaluated by their performance in the downstream recognition or classification problems, i.e., from an \emph{analysis} point of view. 
For example, a timbre-invariant feature is supposed to work better than a non timbre-invariant one for harmonic analysis. 
It remains unclear how timbre and pitch are actually disentangled in the feature representations.

With the recent success in learning disentangled representations for images using deep autoencoders~\cite{konstantinos16nips,liu2018exploring},
we see new opportunities to tackle timbre and pitch disentanglement for music from the \emph{synthesis} point of view.
Taking a musical audio clip as input, we aim to build a model that processes the timbre and pitch information in two different streams to arrive at the intermediate timbre and pitch representations that are disentangled, and that can be combined to \emph{reconstruct} the original input. 
If timbre and pitch are successfully disentangled, we expect that we can change the instrumentation of a music clip 
with this model 
by manipulating the timbre representation only, while fixing the pitch representation.\footnote{If we think about timbre as \emph{tone colors}, this is like \emph{coloring} the music clip in different ways. When there are multiple instruments, the model needs to decide which instrument plays which notes.}

To our best knowledge, this work represents the first attempt to disentangle timber and pitch in music audio with deep encoder/decoder architectures. Our approach has a few advantages over the conventional analysis approach.
First, we adopt a deep neural network to learn features in a data-driven way, instead of hand-crafting the features.
Second, we can use the learned features to generate an audio signal, which provides a direct way to evaluate the effectiveness of disentanglement.
Third, accordingly, it enables new applications in audio-domain \emph{music editing}---to manipulate the timbre and pitch content of an existing music clip without re-recording it.
Lastly, given a note sequence generated by human or an AI composer~\cite{fernandez13jair,dong2017musegan,liu18ismirlbd},  
our model can help decide how to color (i.e., add timbre to) it.

Due to the differences in images and music, we cannot directly apply existing methods to music. 
Instead, we propose two ideas that consider the specific characteristics of music in designing the encoders and decoders.

The first idea is \emph{temporal supervision}.
In computer vision, people use image-level attributes as the supervisory signal to learn disentangled features. For example, to disentangle face identities and poses~\cite{tran2017disentangled}, or to disentangle different attributes of faces~\cite{liu2018exploring}.
For music, we cannot analogously use clip-level labels, since timbre and pitch are associated with each individual musical notes, and a music clip is composed of multiple notes. Therefore, we propose to use the multi-track \emph{pianorolls} \cite{pypianoroll} as the learning target of our encoders/decoders to provide detailed temporal supervision at the frame level.
That is, instead of aiming to reconstruct the input audio, we aim to generate as the output of the network the pianoroll associated with the input audio. 
A pianoroll is a symbolic representation of music that specifies the timbre and pitch per note. It can be derived from a MIDI file~\cite{pypianoroll}.
We manage to compile a new dataset with 350,000 pairs of audio clips and time-aligned MIDIs. 
The temporal supervision provided by the dataset greatly facilitates timbre and pitch disentanglement, for otherwise the model has to learn from the audio signals in an unsupervised way.
We intend to make public the dataset and our code for reproducibility.

The second idea is to use different operations to deal with timbre and pitch. Specifically, we propose to use \emph{convolutions} in the encoder to learn the abstract timbre representation in the latent space, and use \emph{symmetric skip connections} to allow the pitch information to flow directly from the encoder to the decoder. This design is based on the intuition that, when we use a time-frequency representation such as the spectrogram as the input, the timbre aspect actually affects \emph{how} the energy of the harmonic partials distributes along the frequency axis and develops over time \cite{mueller11jstsp}, whereas the pitch aspect determines only \emph{what} the fundamental frequency and duration of the notes are.
To make an analogy between music and images, pitch acts like the boundary of visual objects, whereas timbre acts like the texture. We therefore aim to learn an embedding for the texture, while process the discrete pitch information with skip connections, which have been found effective in image segmentation \cite{ronneberger2015u}.  

\begin{figure}[t]
\centering
\begin{subfigure}[b]{0.5\textwidth} 
\centering
\includegraphics[width=\textwidth]{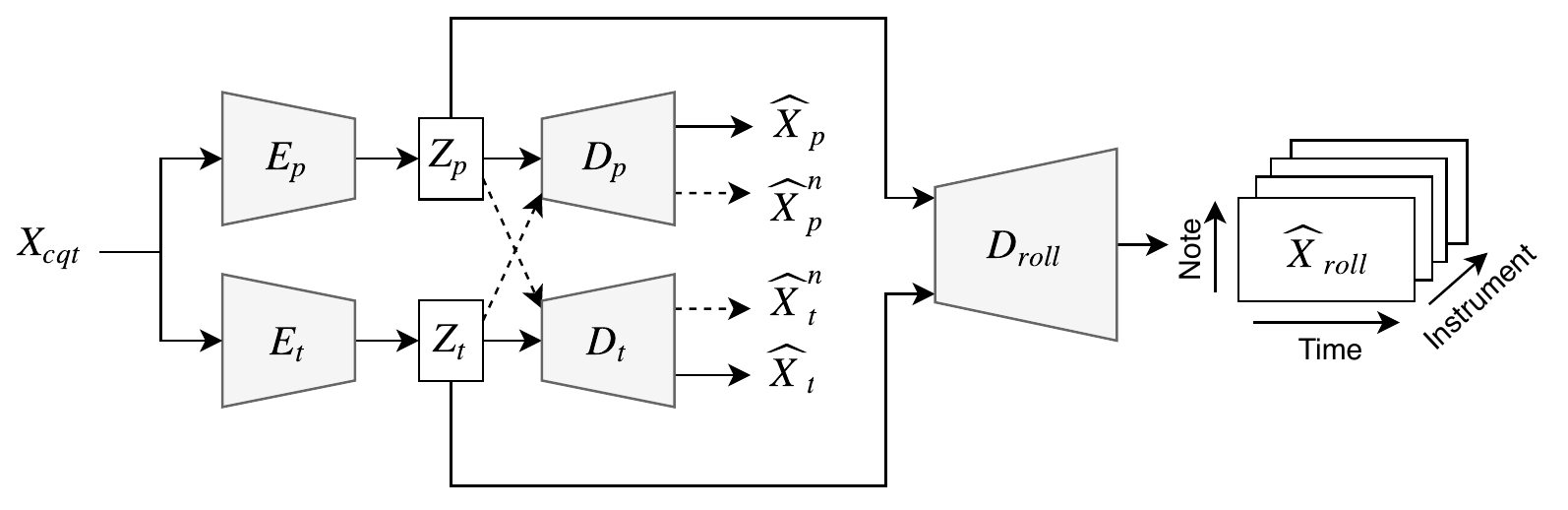}
\caption{DuoAE: Use separate encoders/decoders for timbre and pitch}
\label{fig: xnet}
\vspace{3mm}
\centering
\includegraphics[width=0.8\textwidth]{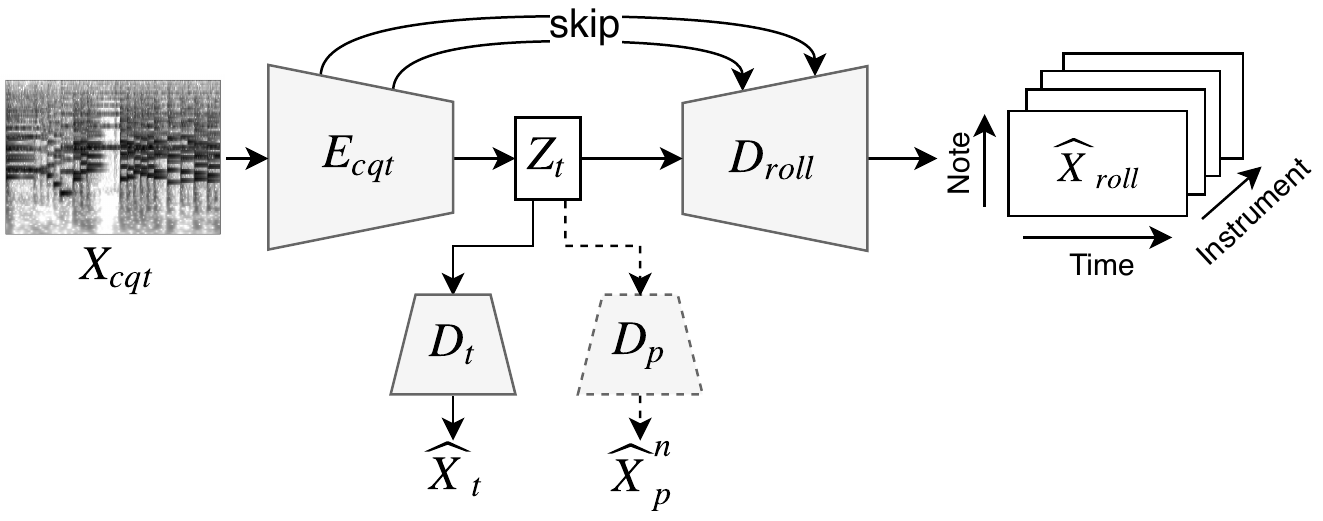}
\caption{UnetAE: Use skip connections to process pitch}
\label{fig: unet}
\end{subfigure}
\caption{The two proposed encoder/decoder architectures for disentangling timbre and  pitch for music audio. The dashed lines indicate the adversarial training parts. (Notations: CQT---a time-frequency representation of audio; roll--multi-track pianoroll; E---encoder; D---decoder; Z---latent code; t---timbre; p---pitch; skip---skip connections).}
\label{fig: model}
\end{figure}

Specifically, we propose two models for disentangling timbre and pitch. The first model (DuoAE) uses separate encoders for timbre and pitch, whereas the second model (UnetAE) adopts the aforementioned second idea and use skip connections to deal with pitch. 
Both models employ adversarial training \cite{mathieu16nips}. 
Figure \ref{fig: model} illustrates the two models. We will present the model details later.


As secondary contributions, for music editing and generation purposes, we additionally train another encoder/decoder sub-network to convert the pianorolls into audio signals. We find that the use of \emph{binary neurons}~\cite{dong2018convolutional} is critical for this new sub-network. In addition, we propose a new evaluation protocol called ``timbre crossover'' to evaluate how well we can create a new music by exchanging the instrumentation of two existing pieces, using the timbre representation of one piece in reconstructing the pianoroll or audio of the other piece.
We report systematic objective evaluations of the result of timbre crossover.

\section{Background}
\label{sec_related_work}


A deep autoencoder (or AE for short) is a network architecture that uses a stack of encoding layers (known collectively as an \emph{encoder}) to get a low-dimensional representation of data, which originally resides in a high-dimensional space~\cite{masci11icann}. The resulting representation is also known as the \emph{latent code}. 
The network is trained such that we can recover the input from the latent code, by passing the latent code through another stack of decoding layers (known collectively as a \emph{decoder}).
Compared to other representation learning methods, the AE has the advantages that the training is unsupervised, and that the obtained representation can be mapped back to the data space.

With the original AE, different properties of data might be entangled 
in the latent code, meaning that each entry of the latent code is related to multiple data properties.
For instance, we can train an AE with face images to obtain a latent vector $\mathbf{z} \in \mathbb{R}^k$ for an image, where $k$ denotes the dimensionality of the latent code. If we add some noise $\epsilon$ to only a random entry of $\mathbf{z}$ and then decode it, likely many properties of the decoded face 
would be different from the original face.

With some labeled data, we can train an AE in a way that different parts of the latent code correspond to different data properties.
For example, given face images labeled with identities and poses, we can divide $\mathbf{z}$ into two parts $\mathbf{z}_1$ and $\mathbf{z}_2$ (i.e., $\mathbf{z}=[\mathbf{z}_1^T, \mathbf{z}_2^T]^T$), and use them as the input to two separate stacks of fully-connected layers to train an identity classifier $C_1$ and a pose classifier $C_2$, respectively. We still require the concatenated code (i.e., $\mathbf{z}$) to reconstruct the input. With this AE, if we 
add changes to $\mathbf{z}_2$ but keep $\mathbf{z}_1$ the same, 
we may obtain an image with the same identity but a different pose. In this way, we call $\mathbf{z}_1$ and $\mathbf{z}_2$  disentangled representations of face identities and poses.

We can further improve the result of disentanglement by using \emph{adversarial training}. As usual, we aim to minimize the classification error of $C_1$ and $C_2$ when their input is $\mathbf{z}_1$ and $\mathbf{z}_2$, respectively. However, we 
additionally use $\mathbf{z}_2$ and $\mathbf{z}_1$ as the input to $C_1$ and $C_2$ respectively, and aim to \emph{maximize} the classification error of the two classifiers under such a scenario. In this way, we promote identity information in $\mathbf{z}_1$ while \emph{dispel} any information related to pose, and similarly for $\mathbf{z}_2$. This idea was proposed by~\cite{liu2018exploring}.

There are many other ways to achieve disentanglement, e.g., using generative adversarial networks (GAN)~\cite{tran2017disentangled,huang2017beyond,liu2018multi},  
cross-covariance penalties~\cite{cheung2014discovering},
and latent space arithmetic operations~\cite{hsu2017learning}. Some are unsupervised methods.
While the majority of work has been on images, feature disentanglement has also been studied for video clips~\cite{hsieh2018learning}, speech clips~\cite{hsu2017learning}, 
3D data \cite{liu2018disentangling}, and MIDIs~\cite{roberts2018hierarchical,brunner2018midi}. 

Our work distinguishes itself from the existing works mainly in the following two aspects. First, we use temporal supervision to learn disentangled representations, while existing work usually use image- or clip-level labels, such as face identity, speaker identity~\cite{hsu2017learning}, and genre~\cite{brunner2018midi}. 
Second, to our best knowledge, little has been done for disentangled representation learning in music audio.
Although MIDI is also a type of music, an disentangling model for encoding MIDIs cannot be applied to musical audio signals.\footnote{For example, the sounds of different partials from different instruments are mixed in audio, but this does not happen in MIDIs.}\footnote{Disentangling musical properties in MIDIs is a challenging task as well.  
\cite{roberts2018hierarchical} aimed to disentangle melody and instrumentation using a variational autoencoder (VAE), but they found that manipulating the instrumentation code (while fixing the melody code) would still change the melody.}

\section{Proposed Models for Disentanglement}
\label{sec_method}

Figure \ref{fig: model} shows the architecture of the proposed models for disentangling timbre and pitch. We present the details below.


\subsection{Input/Output Data Representation}

\subsubsection{Input}
The input to our models is an audio waveform with arbitrary length. To facilitate timbre and pitch analysis, we firstly convert the waveform into a time-frequency representation that shows the energy distribution across different frequency bins for each short-time frame. Instead of using the short-time Fourier transform (STFT), we use the constant-Q transform (CQT) here, for the latter adopts a logarithmic frequency scale that better aligns with our perception of pitch~\cite{bittner17ismir}. CQT also provides better frequency resolution in the low-frequency part, which helps detect the fundamental frequencies.

As will be shown later, our encoders and decoders are designed to be \emph{fully-convolutional}~\cite{oquab15localization}, so that our models can deal with input of any length in testing time. However, for the convenience of training the models with mini-batches, in the training stage we divide the waveforms in our training set into 10-second chunks (without overlaps) and use these chunks as the model input, leading to a matrix $\mathbf{X}_{cqt} \in \mathcal{R}^{F \times T}$ of fixed size for each input.
In our implementation, we compute CQT with the \texttt{librosa} library~\cite{librosa}, with 16,000 Hz sampling rate and 512-sample window size, again with no overlaps. We use a frequency scale of 88 bins, with 12 bins per octave to represent each note. Hence, $F=88$ (bins) and $T=312$ (frames).

\begin{figure}
\centering
\begin{subfigure}[b]{0.17\textwidth}
\includegraphics[width=\textwidth]{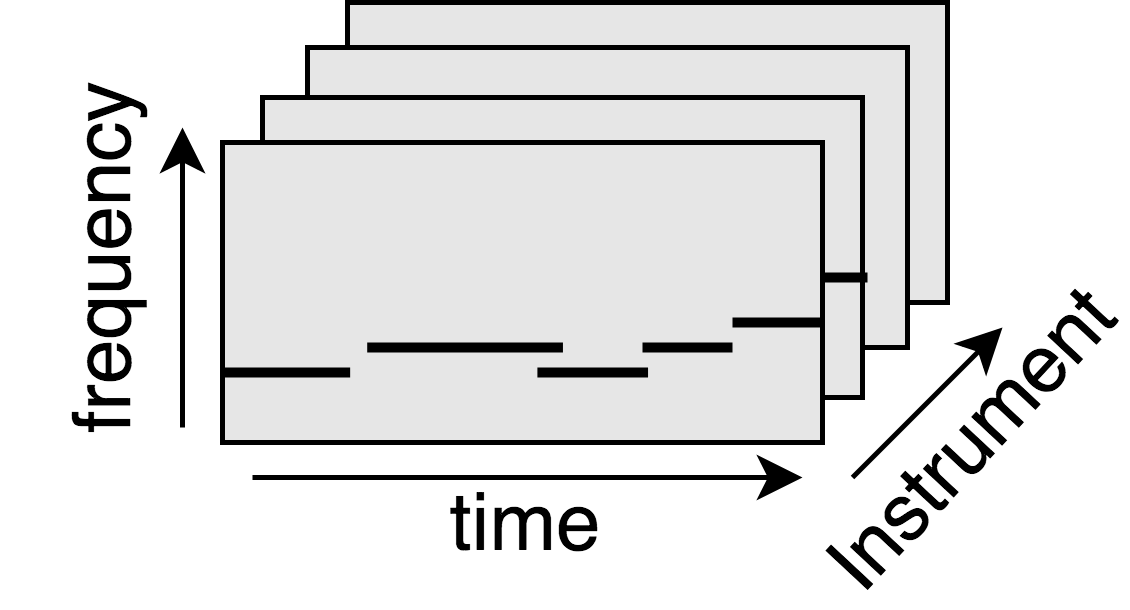}
\caption{Pianoroll}
\end{subfigure}
\begin{subfigure}[b]{0.14\textwidth}
\includegraphics[width=\textwidth]{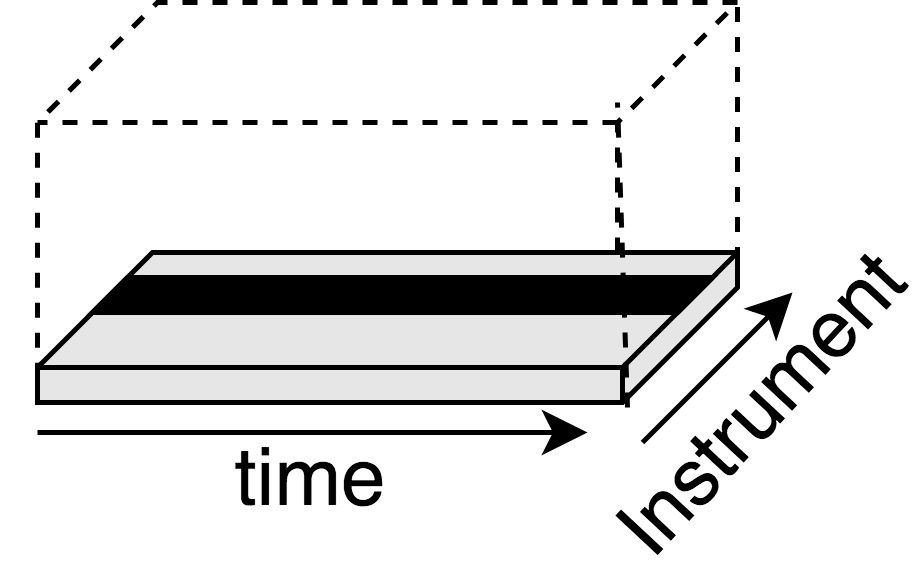}
\caption{Instrument roll}
\end{subfigure}
\begin{subfigure}[b]{0.14\textwidth}
\includegraphics[width=\textwidth]{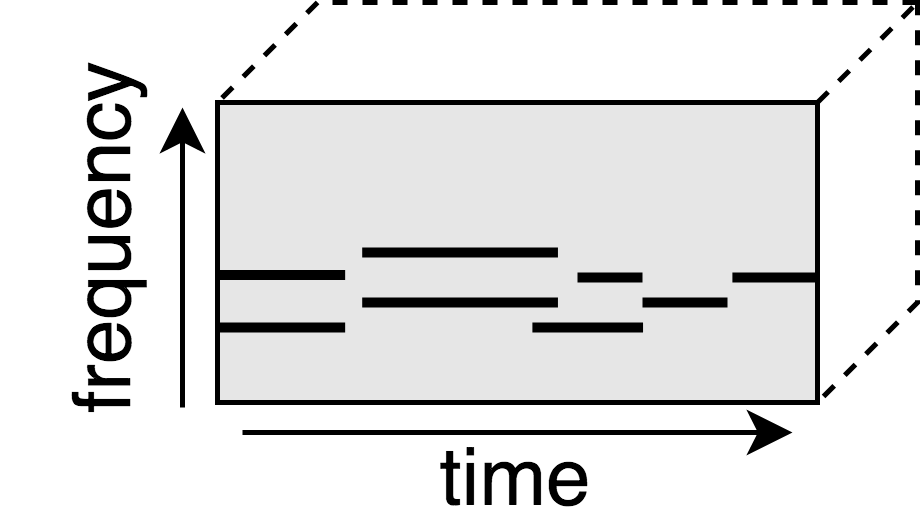}
\caption{Pitch roll}
\end{subfigure}
\caption{Different symbolic representations of music.}
\label{fig_pianoroll}
\end{figure}

\subsubsection{Output}
To provide temporal supervision, we use the pianorolls
as the target output of our models.\footnote{Strictly speaking, such a model is no longer an autoencoder, since the input and target output are different. We abuse the terminology for the short names of our models, e.g., Duo`AE.'}
As depicted in Figure \ref{fig_pianoroll}, a pianoroll is a binary-valued tensor that records the presence of notes (88 notes here) across time for each track (i.e., instrument)  \cite{pypianoroll}. When we consider $M$ instruments, the target model output would be $\mathbf{X}_{roll} \in \{0,1\}^{F \times T \times M}$. 
$\mathbf{X}_{roll}$ and $\mathbf{X}_{cqt}$ are temporally aligned, since we use MIDIs that are time-aligned with the audio clips to derive the pianorolls, as will be discussed in Section \ref{sec_data}.

As shown in Figure \ref{fig: model}, besides asking our models to generate $\mathbf{X}_{roll}$ from the latent code of $\mathbf{X}_{cqt}$, we use the \emph{instrument roll} $\mathbf{X}_{t} \in \{0,1\}^{M \times T}$ and \emph{pitch roll} $\mathbf{X}_{p} \in \{0,1\}^{F \times T}$ as supervisory signals to disentangle timbre and pitch. As depicted in Figure~\ref{fig_pianoroll}, these two rolls can be obtained respectively by marginalizing a certain dimension of the pianoroll. 




\subsection{The DuoAE Model}

The architecture of DuoAE is illustrated in Figure \ref{fig: model}(a). The designed is based on~\cite{liu2018exploring}, but we use temporal supervision here and adapt the model to encode music.

Specifically, we train two encoders $E_t$ and $E_p$ to respectively convert $\mathbf{X}_{cqt}$ into the \emph{timbre code} $\mathbf{Z}_t = E_t(\mathbf{X}_{cqt}) \in \mathcal{R}^{\kappa \times \tau}$ and \emph{pitch code} $\mathbf{Z}_p = E_p(\mathbf{X}_{cqt}) \in \mathcal{R}^{\kappa \times \tau}$.
We note that, unlike in the case of image representation learning, here the latent codes are \emph{matrices}, and we require that the second dimensions (i.e., $\tau$) represent time. This way, each column of $\mathbf{Z}_t$ and $\mathbf{Z}_p$ is a $\kappa$-dimensional representation of a temporal segment of the input. For abstraction, we require $\kappa\tau < FT$.

DuoAE also contains three decoders $D_{roll}$, $D_t$ and $D_p$.  The encoders and decoders are trained such that we can use $D_{roll}([\mathbf{Z}_t^T, \mathbf{Z}_p^T]^T)$ to predict $\mathbf{X}_{roll}$,  $D_t(\mathbf{Z}_t)$ to predict $\mathbf{X}_t$, and  $D_p(\mathbf{Z}_p)$ to predict $\mathbf{X}_p$. The prediction error is measured by the \emph{cross entropy} between the ground truth and the predicted one. For example, for the timbre classifier $D_t$, it is:
\begin{equation} 
L_t = - \textstyle{\sum} ~[\mathbf{X}_t\cdot \ln\sigma(\widehat{\mathbf{X}}_t) + (1-\mathbf{X}_t) \cdot \ln(1-\sigma(\widehat{\mathbf{X}}_t)) ] \,,
\label{eq:lt}
\end{equation}
where $\widehat{\mathbf{X}}_t = D_t(\mathbf{Z}_t)$, 
`$\cdot$' denotes the element-wise product, and 
$\sigma$ is the sigmoid function that scales its input to $[0,1]$. We can similarly define $L_{roll}$ and $L_p$.

In each training epoch, we optimize \emph{both} the encoders and decoders by minimizing $L_{roll}$, $L_t$ and $L_p$ for the given training batch.
We refer to the way we train the model as using the temporal supervision, since to minimize the lost terms $L_{roll}$, $L_t$ and $L_p$, we have to make accurate prediction for each of the $T$ time frames.



When the adversarial training strategy is employed (i.e., those marked by dashed lines in Figure \ref{fig: model}(a)), we additionally consider the following two lost terms:
\begin{equation} 
L_t^n = - \textstyle{\sum} ~[\mathbf{0}_t\cdot\ln \sigma(\widehat{\mathbf{X}}_t^n) + (1-\mathbf{0}_t)\cdot\ln(1-\sigma(\widehat{\mathbf{X}}_t^n)) ] \,, 
\label{eq:ltn}
\end{equation}
\begin{equation} 
L_p^n =  - \textstyle{\sum} ~[\mathbf{0}_p\cdot\ln \sigma(\widehat{\mathbf{X}}_p^n) + (1-\mathbf{0}_p)\cdot\ln(1-\sigma(\widehat{\mathbf{X}}_p^n)) ] \,, 
\label{eq:lpn}
\end{equation}
where $\widehat{\mathbf{X}}_t^n = D_t(\mathbf{Z}_p)$, $\widehat{\mathbf{X}}_p^n = D_p(\mathbf{Z}_t)$, meaning that we feed the `wrong' input (purposefully) to $D_t$ and $D_p$.
Moreover, $\mathbf{0}_{t}=\mathbf{0}_{M,T}$ and $\mathbf{0}_{p}=\mathbf{0}_{F,T}$ are two matrices of \emph{all zeros}. 
That is to say, when we use the wrong input, we expect $D_t$ and $D_p$ can output nothing (i.e., all zeros), since, e.g., $\mathbf{Z}_p$ is supposed not to contain any timbre-related information.

Please note that, in adversarial training, we use $L_t^n$ and $L_p^n$ to update the \emph{encoders only}. This is to preserve the function of the decoders in making accurate predictions.

\subsection{The UnetAE Model}
The architecture of UnetAE is depicted in Figure \ref{fig: model}(b).\footnote{We explain the name `Unet' below. When the design of the encoder (E) and decoder (D) is \emph{symmetric}, meaning that they have the same number of layers and that they use the same kernel sizes and stride sizes in the corresponding layers, we can add skip connections between the corresponding layers of E and D, by concatenating the output of the $i$-th layer of E to the input of the $i$-th last layer of D in the channel-wise direction. In this way, lower-layer information of E (closer to the input) can be directly passed to the higher-layer of D (closer to the output), making it easier to train deeper AEs. Because the resulting architecture has a U-shape, people refer to it as a U-net~\cite{ronneberger2015u}.}
In UnetAE, we learn only one encoder $E_{cqt}$ to get a single latent representation $\mathbf{Z}_t$ of the input $\mathbf{X}_{cqt}$. 
We add skip connections between $E_{cqt}$ and $D_{roll}$ and 
learn $E_{cqt}$ and $D_{roll}$ by minimizing $L_{roll}$, the cross entropy between $D_{roll}(\mathbf{Z}_t)$ and the pianoroll $\mathbf{X}_{roll}$. Moreover, we promote timbre information in $\mathbf{Z}_t$ by refining $E_{cqt}$ and learning a classifier $D_t$ by minimizing $L_t$ (see Eq.~(\ref{eq:lt})). When the adversarial training strategy is adopted, we use the classifier $D_p$ pre-trained from DuoAE to further dispel pitch information from $\mathbf{Z}_t$, by updating $E_{cqt}$ (but fixing $D_p$) to minimize $L_p^n$ (see Eq.~(\ref{eq:lpn})).

In summary, we use $L_{roll}$, $L_t$, and optionally $L_p^n$, to train the encoder $E_{cqt}$; use $L_{roll}$ to train the decoder $D_{roll}$; and use $L_t$ to train the timbre classifier $D_t$. 
We discard the lost term $L_p^n$ when we do not use adversarial training.

The two key design principals of UnetAE are as follows. First, since $\mathbf{Z}_t$ is supposed not to have any pitch information, the only way to obtain the pitch information needed to predict $\mathbf{X}_{roll}$ is from the skip connections. 
Second, there are reasons to believe that the skip connections can pass along the pitch information, because there is nice one-to-one time-frequency correspondence between $\mathbf{X}_{cqt}$ and each frontal slice of $\mathbf{X}_{roll}$,\footnote{That is, both $X_{cqt}(i,j)$ and $X_{roll}(i,j,m)$ refer to the activity of the same musical note $i$ for the same time frame $j$.} and because in $\mathbf{X}_{cqt}$ pitch only affects the lowest partial of a harmonic series created by a musical note, while timbre affects all the partials. If we view pitch as the boundary outlining an object (i.e., the harmonic series) and timbre as the texture of that object, it makes sense to use a U-net structure, since U-net performs well in image segmentation~\cite{badrinarayanan17tpami,ronneberger2015u}.
%



\begin{figure*}[t]
\centering
\includegraphics[width=.75\linewidth]{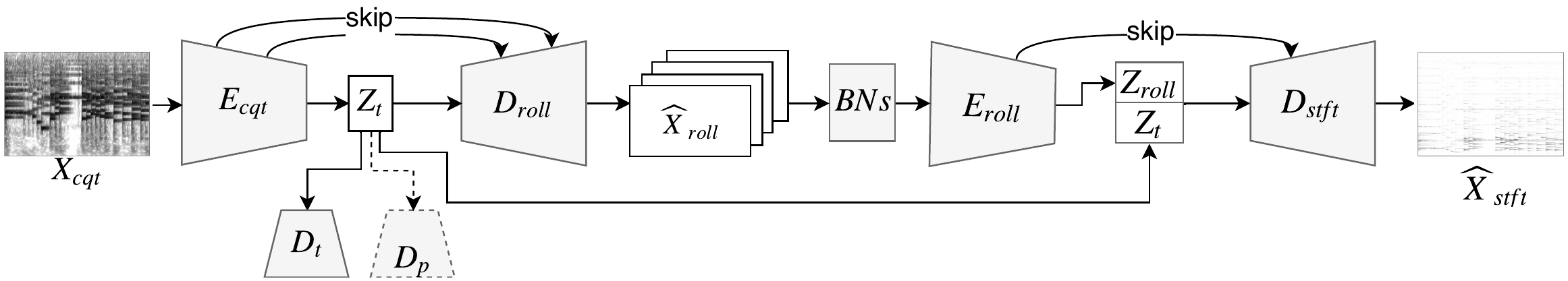}
    \caption{The architecture of the proposed model for music audio editing, using UnetAE for disentangling timbre and pitch.}
    \label{fig: editing}
\end{figure*}

\subsubsection{Discussion}
The major difference between DuoAE and UnetAE is that there is no pitch code $\mathbf{Z}_p$ in UnetAE. We argue below why this may be fine for music editing and generation applications.

As discussed in the introduction, pitch determines \emph{what} (i.e., the notes) to be played in a music piece, whereas timbre determines \emph{how} they would sound like to a listener. While the pitch content of a music piece can be largely specified on the musical score (i.e., symbolic notations of music), the timbre content manifests itself in the audio sounds. Therefore, we can learn the timbre code $\mathbf{Z}_t$ from an input audio representation such as the waveform, STFT, or CQT. If we want to learn the pitch code $\mathbf{Z}_p$, we can learn it from symbolic representations of music, as pursued in existing work on symbolic-domain music generation~\cite{yang2017midinet,roberts2018hierarchical,brunner2018midi,simon18ismir,dong2017musegan}. 
Since the pitch code can be learned from musical scores, we may focus on learning the timbre code from music audio.

Moreover, for music audio editing, we are interested in manipulating the instrumentation of a clip without much affecting its pitch content. For this purpose, it may be sufficient to have the timbre code, since in UnetAE the pitch content can flow directly from the encoder to the decoder. On the other hand, to manipulate the pitch content without affecting the timbre, we can change the pitch in the symbolic domain and then use the same timbre code for decoding.

\subsubsection{Implementation Details}

Since the input and target output are both matrices (or tensors), we use convolutional layers in all the encoders and decoders of DuoAE and UnetAE. To accommodate input of variable length, we adopt a \emph{fully-convolutional} design~\cite{oquab15localization}, meaning that we do not use pooling layers at all. In the encoders, we achieve dimension reduction (i.e., reducing $F$ to $\kappa$ and reducing $T$ to $\tau$) by setting the stride sizes of the kernels larger than one. We use $3\times 3$ kernels in every layers of the encoders. In the decoders, we use transposed convolution for upsampling. The kernel size is also $3\times3$. 
Moreover, we use leaky ReLU as the activation function and add batch normalization to all but the last layer of the decoders, where we use the sigmoid function. 
Both DuoAE and UnetAE are trained using stochastic gradient descend (SGD) with momentum 0.9. The initial learning rate is set to 0.01.

\section{Proposed Model for Music Editing}

The model we propose for music editing is shown in Figure \ref{fig: editing}. We use an additional encoder-decoder subnet $E_{roll}$ and $D_{stft}$ after UnetAE to convert pianorolls to the audio domain.\footnote{Such a pianoroll-to-audio conversion can be made with commercial synthesizers, but the resulting sounds tend to be deadpan.} Here we use the STFT spectrograms of the original input audio as the target output. 
The model is trained in an end-to-end fashion by additionally minimizing $ L_{MSE}$, the MSE between the groundtruth STFT $\mathbf{X}_{stft}$ and the predicted one $\widehat{\mathbf{X}}_{stft}$. 
Lastly, we use the Griffin-Lim algorithm \cite{griffin84} to estimate the phase of the spectorgram and generate the audio.

Once the model is trained, timbre editing can be done by manipulating the timbre code $\mathbf{Z}_t$ and then generate $\widehat{\mathbf{X}}_{stft}$. And, pitch editing can be done by manipulating the intermediate pianoroll $\widehat{\mathbf{X}}_{roll}$ and then generate $\widehat{\mathbf{X}}_{stft}$.

There are three critical designs to make it work.
First, we concatenate $\mathbf{Z}_t$ with $\mathbf{Z}_{roll}$, the output of $E_{roll}$. This is important since the input to $E_{roll}$ is pianorolls, but the pianorolls do not contain sufficient timbre information to generate realistic audio.
Second, we only use one skip connection between $E_{roll}$ and $D_{stft}$, because if we use too many skip connections the gradients from the STFT will affect the training of $D_{roll}$ and lead to noisy pianorolls $\widehat{\mathbf{X}}_{roll}$.
Third, to further prevent the gradients from the STFT to affect the pianorolls, 
we use deterministic \emph{binary neurons} (BN) \cite{dong2018convolutional} 
to binarize the output of $D_{roll}$
with a hard thresholding function.

\subsubsection{Discussion}

A few other models have been proposed to generate musical audio signals.
Many of them do not take auxiliary conditional signal to condition the generation process. These include auto-regressive models such as the WaveNet model \cite{van2016wavenet} 
and GAN-based models such as the WaveGAN model \cite{wavegan}.
Some recent works started to explore the so-called \emph{score-to-audio} music generation \cite{wang19aaai,hawthorne2018enabling}, where the audio generation model is given a musical score and is asked to render the score into sounds. 
While the model proposed by Hawthorne \emph{et al.} \cite{hawthorne2018enabling} deals with only piano music, both our music editing model and the PerformanceNet model proposed by Wang and Yang \cite{wang19aaai} aim to generate music of more other instruments. 

However, our model is different from these two prior arts in that we are not dealing with score-to-audio music generation but actually audio editing, which converts an audio into another audio. 
We use the estimated pianoroll $\widehat{\mathbf{X}}_{roll}$ in the midway of the generation process for controllability. 
As Figure \ref{fig: editing} shows, our model takes not only the pianoroll $\widehat{\mathbf{X}}_{roll}$ but also the timbre code $\mathbf{Z}_t$ of the original audio as inputs. We find such a problem setting interesting and would try to further improve it in our future work, using for example WaveNet-based decoder or GAN training for better audio quality.


\begin{table*}
\centering
\begin{tabular}{|l|c|c|cc|cc|cc|} 
\hline
\multirow{2}{*}{Method} &\multirow{2}{*}{\#Params}& Transcription 
&\multicolumn{2}{c}{Crossover~(*$\rightarrow$piano)} 
& \multicolumn{2}{c}{Crossover~(*$\rightarrow$guitar)}
& \multicolumn{2}{c|}{Crossover~(pia$\rightarrow$str)}\\
\cline{3-9}
& & Acc&Pitch Acc&Timbre HI&Pitch Acc&Timbre HI &Pitch Acc&Timbre HI \\
\hline
Baseline model	&3,277k &0.314&---&---&---&---&---&--- \\ 
Prior art  \cite{liu2018multi}  &6,533k&---&0.558&---&0.537&---&---&---\\
Prior art \cite{hadad2018two}  &7,729k&---&0.295&---&0.279&---&---&---\\
\hline
DuoAE~w/o~adv   &9,438k &0.395&0.595&0.990&0.664&0.983&0.743&0.998 \\
DuoAE 			&9,808k &0.372&0.660&0.990&0.647&0.989&0.718&0.998 \\
UnetAE~w/o~adv  &3,499k &0.396&0.563&0.869&0.611&0.911&0.681&0.966 \\
UnetAE 			&3,868k &0.431&\textbf{0.691}&0.893&\textbf{0.727}&0.962&0.748&0.992 \\
\hline
\end{tabular}
\caption{The performance of different models for transcription (i.e., pianoroll prediction) and timbre crossover, evaluated in terms of pitch accuracy (Acc) and the timbre histogram intersection (HI) rate. 
The last two columns are `piano to violin+cello' conversion. We use `w/o adv' to denote the cases without adversarial training, and `\#Params' the total number of parameters. The `baseline model' (see Section \ref{sec:exp_1}) shown in the first row does not use timbre and pitch classifiers and adversarial training. }
\label{tab:pianorolls Acc}
\end{table*}

\section{Dataset}
\label{sec_data}
 
We build a new dataset with paired audio and MIDI files to train and evaluate the proposed models. 
This is done by crawling the MuseScore web forum (\url{https://musescore.com/}), obtaining around 350,000 unique MIDI files and the corresponding MP3 files. Most MP3 files were synthesized from the MIDIs with the MuseScore synthesizer by the uploaders.\footnote{Though synthesized audio may sounds different than realistic audio, the problem can be solved by domain adaptation \cite{konstantinos16nips}. We take it as a future task and will not further discussed in this paper. Besides, we also tested the model on realistic music and found that the model performs well for many of them.} 
Hence, the audio and MIDIs are already time-aligned.
We further ensure temporal alignment by using the method proposed by \cite{raffel2016optimizing}.
We then convert the time-aligned MIDIs to pianorolls with the \texttt{Pypianoroll} package~\cite{pypianoroll}. 

We consider the following nine instruments in this work (i.e., $M=9$): piano, acoustic guitar, electrical guitar, trumpet, saxphone, bass, violin, cello and flute. They are chosen based on their popularity in modern music and MuseScore.\footnote{We exclude drums for they are non-pitched. And, we exclude the singing voices, for they are not transcribed in MIDIs.} 
The average length of the music clips is $\sim$2 minutes. 
We randomly pick 585--1000 clips per instrument for the training set, and 27--50 clips per instrument for the test set. 
As a result, the training and test sets are class-balanced.



\section{Experiment}

In what follows, we first evaluate the accuracy of our models in predicting the pianorolls. We then evaluate the result of timbre and pitch disentanglement by examining the learned embeddings and by the timbre crossover evaluation method.

\subsection{Evaluation on Pianoroll Prediction}
\label{sec:exp_1}

We first evaluate how well we can transcribe the pianorolls from audio.  Since $\widehat{\mathbf{X}}_{roll}$, the
output of $D_{roll}$, is a real-valued tensor in $[0,1]^{F \times T \times M}$,
we further binarize it with a simple threshold picking algorithm  so that we can compare it with the groundtruth pianoroll $\mathbf{X}_{roll}$, which is binary. We select the threshold (from 0.1, 0.15, $\dots$, 0.95, in total 20 candidates) by maximizing the accuracy on a held-out validation set.
The accuracy (denoted as `Acc') is calculated by comparing $\mathbf{X}_{roll}$ and $\widehat{\mathbf{X}}_{roll}$ per instrument (by calculating the proportion of true positives among the $FT$ entires) and then taking the average across the instruments. 
This way, we can measure the accuracy for both instrument and pitch prediction, since falsely predicting the instrument of a note would also reduce the number of true positives for instruments.

In addition to DuoAE and UnetAE, we consider a \textbf{baseline model} that uses only an encoder $E_{cqt}$ and a decoder $D_{roll}$ to get the latent code $\mathbf{Z}_t$. That is, not using additional timbre and pitch classifiers and adversarial training. 

The first three columns of Table \ref{tab:pianorolls Acc} show the result. 
Both DuoAE and UnetAE perform much better than the baseline. 
In addition, UnetAE outperforms DuoAE, despite that UnetAE uses much fewer parameters than DuoAE.
We attribute this to the skip connections, which bring detailed information from the input to the decoding process and thereby help pitch localization. The best accuracy 0.431 is achieved by UnetAE, with adversarial training.

\begin{figure}[h!]
\centering
\includegraphics[width=\linewidth]{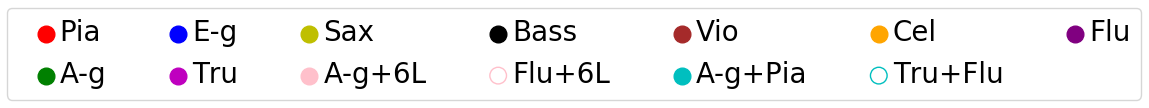}
\begin{subfigure}[c]{\linewidth}
\includegraphics[width=0.46\textwidth]{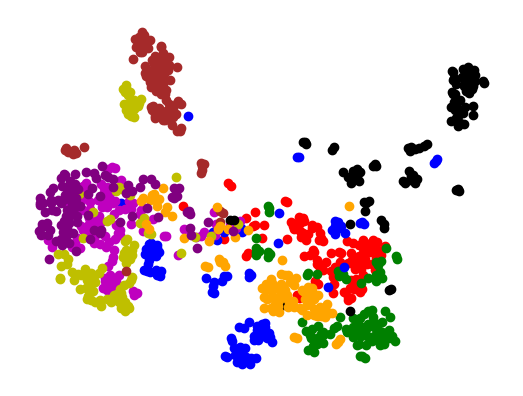}
\includegraphics[width=0.46\textwidth]{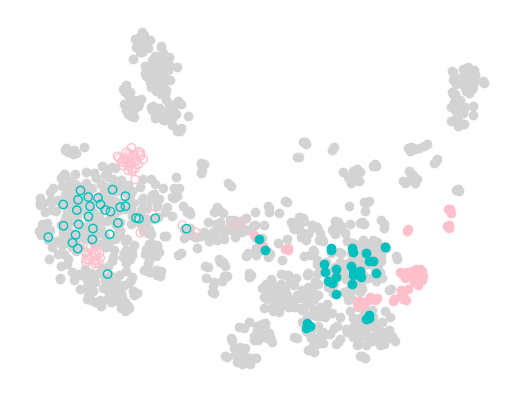}
\caption{Baseline model}
\label{fig: tsne-baseline}
\end{subfigure}
\begin{subfigure}[c]{\linewidth}
\includegraphics[width=0.46\textwidth]{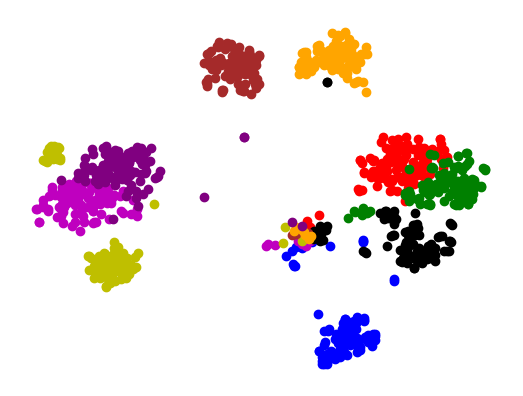}
\includegraphics[width=0.46\textwidth]{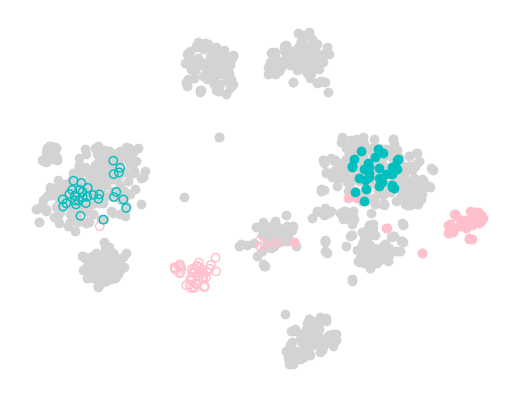}
\caption{DuoAE}
\label{fig: tsne-DuoAE}
\end{subfigure}
\begin{subfigure}[c]{\linewidth}
\includegraphics[width=0.46\textwidth]{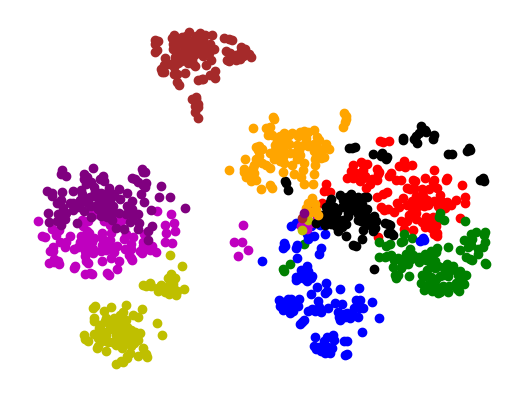}
\includegraphics[width=0.46\textwidth]{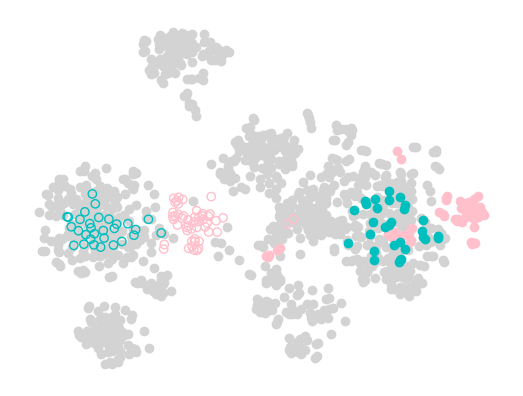}
\caption{UnetAE}
\label{fig: tsne-UnetAEinter}
\end{subfigure}
\caption{t-SNE visualization~\cite{maaten2008visualizing} of the timbre code learned by different models (best viewed in color). Left: the timbre codes of instrument solos. Right: the timbre code of audio chunks with two instruments, or of manipulated solo chunks that are six semitones lower than the original (`6L'). The instruments are piano, electric guitar, saxphone, bass, violin, cello, flute, acoustic guitar, trumpet.}
\vspace{-3mm}
\label{fig: tsne}
\end{figure}

\subsection{Evaluation on Disentanglement}
\subsubsection{t-SNE visualization of the learned timbre code} 
The first thing we do to evaluate the performance of timbre and pitch disentanglement is via examining the timbre code $\mathbf{Z}_t$, by 
projecting them from the $\kappa$-dimensional space to a 2-D space with 
distributed stochastic neighbor embedding (t-SNE)~\cite{maaten2008visualizing}. 
We implement the t-SNE algorithm via Sklearn library with learning rate 20, perplexity 30 and \#iteration 1,000. 
The first column of Figure \ref{fig: tsne} shows the learned timbre code for audio chunks of instrument solos randomly picked from MuseScore.
We can see clear clusters of points from the result of DuoAE and UnetAE, which is favorable since an instrument solo involves only one instrument.

The second column of Figure \ref{fig: tsne} shows the learned timbre code for two more cases: 1) audio chunks with two instruments, picked from MuseScore; 2) audio chunks of instrument solos, but with pitch purposefully shifted lower by us. 
For the first case (in cyan), both DuoAE and UnetAE can nicely position the chunks in the middle of the two clusters of involved instruments.
For the second case (in pink), both DuoAE and UnetAE fail to position the chunks within the clusters of the involved instruments, suggesting that the learned timbre code is not perfectly pitch invariant. But, from the distance between the chunks and the corresponding clusters, it seems UnetAE performs slightly better.

\begin{figure*}[!htp]
\centering


\begin{subfigure}[c]{0.16\linewidth}
\includegraphics[width=\linewidth]{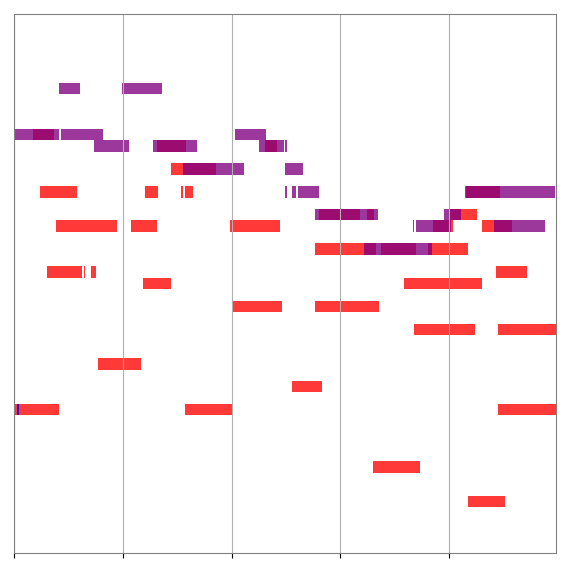}
\caption{\fbox{{pia$+$flu}}}
\label{fig: rolla}
\end{subfigure}
\begin{subfigure}[c]{0.16\linewidth}
\includegraphics[width=\linewidth]{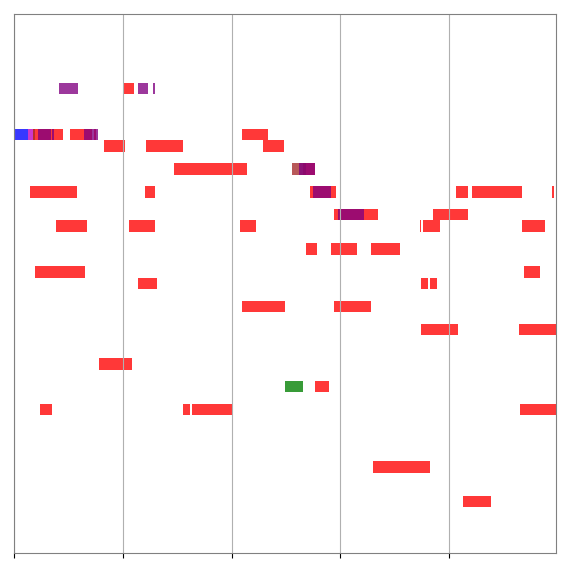}
\caption{$\rightarrow$pia}
\label{fig: rollb}
\end{subfigure}
\begin{subfigure}[c]{0.16\linewidth}
\includegraphics[width=\linewidth]{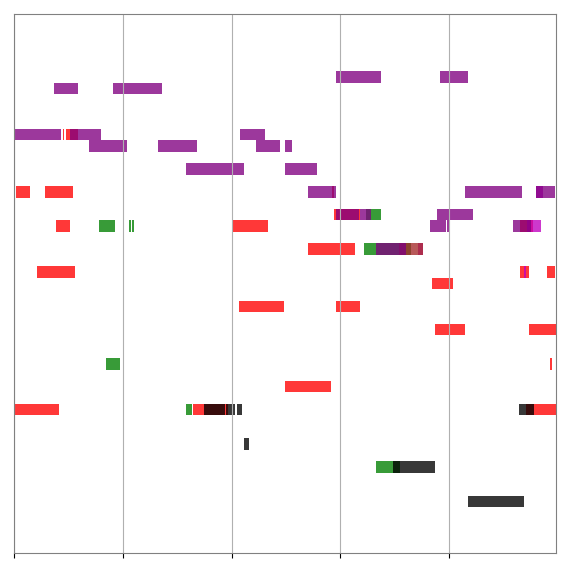}
\caption{$\rightarrow$pia$+$flu$+$bas}
\label{fig: rollc}
\end{subfigure}
\begin{subfigure}[c]{0.16\linewidth}
\includegraphics[width=\linewidth]{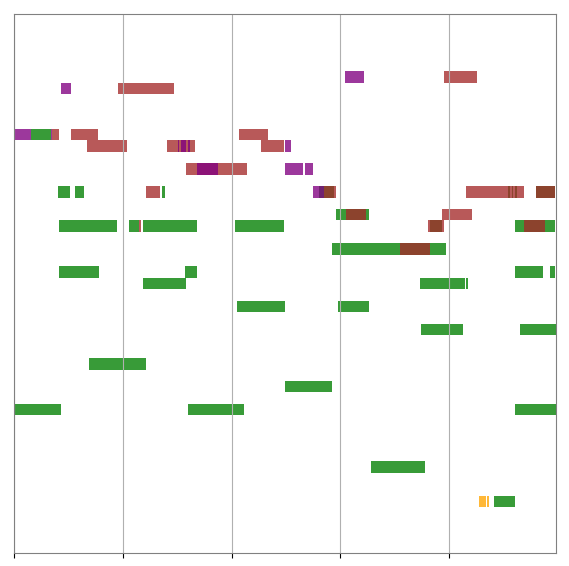}
\caption{$\rightarrow$vio$+$a-g}
\label{fig: rolld}
\end{subfigure}~~~
\begin{subfigure}[c]{0.16\linewidth}
\includegraphics[width=\linewidth]{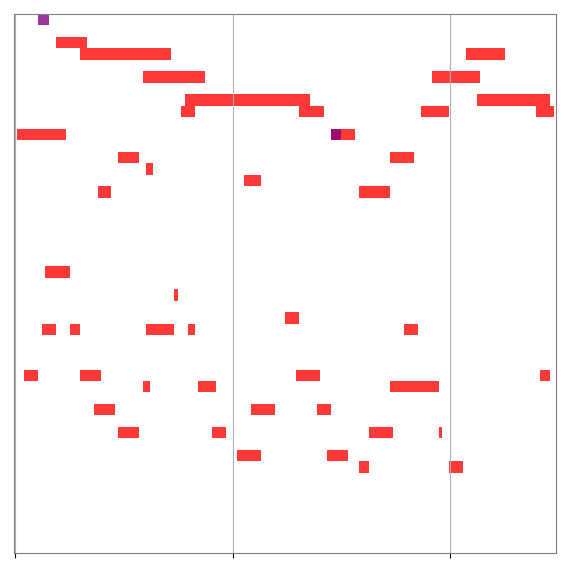}
\caption{\fbox{{pia}}}
\label{fig: rolle}
\end{subfigure}
\begin{subfigure}[c]{0.16\linewidth}
\includegraphics[width=\linewidth]{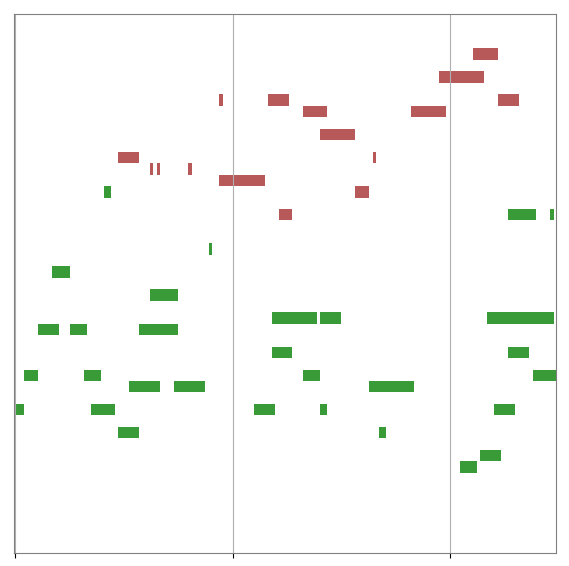}
\caption{$\rightarrow$vio$+$a-g}
\label{fig: rollf}
\end{subfigure}

\begin{subfigure}[c]{0.16\linewidth}
\includegraphics[width=\linewidth]{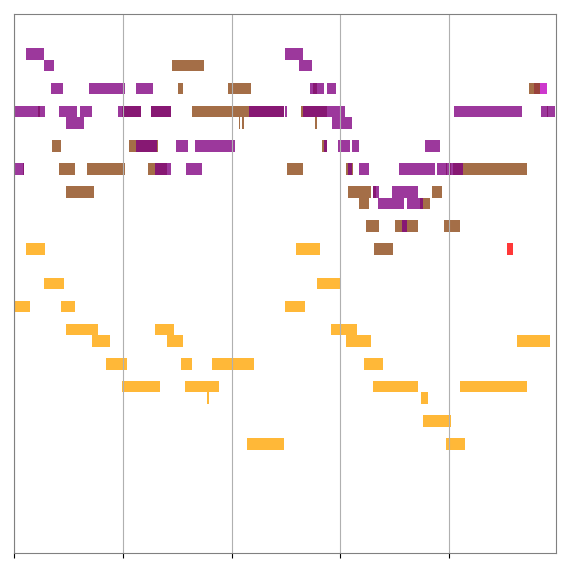}
\caption{\fbox{{vio$+$cel$+$flu}}}
\label{fig: rollg}
\end{subfigure}
\begin{subfigure}[c]{0.16\linewidth}
\includegraphics[width=\linewidth]{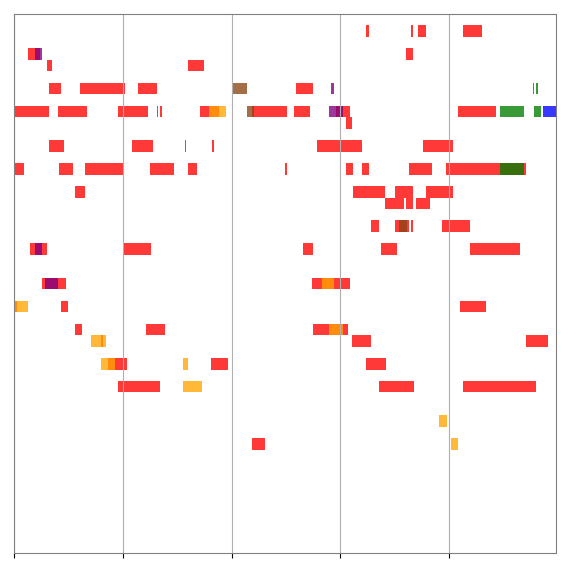}
\caption{$\rightarrow$pia}
\label{fig: rollh}
\end{subfigure}
\begin{subfigure}[c]{0.16\linewidth}
\includegraphics[width=\linewidth]{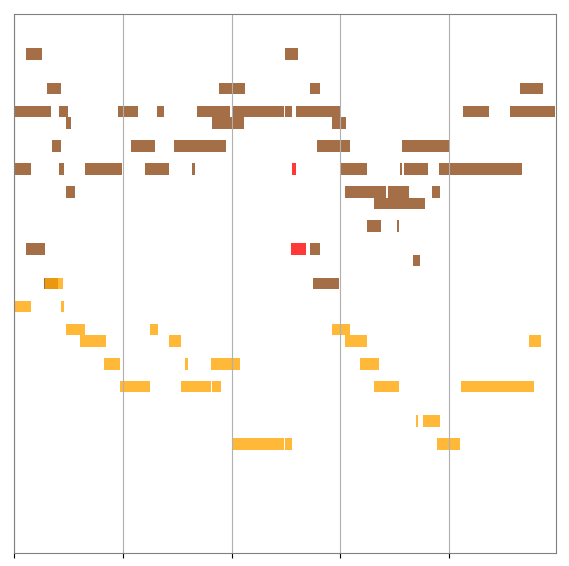}
\caption{$\rightarrow$vio$+$cel}
\label{fig: rolli}
\end{subfigure}
\begin{subfigure}[c]{0.16\linewidth}
\includegraphics[width=\linewidth]{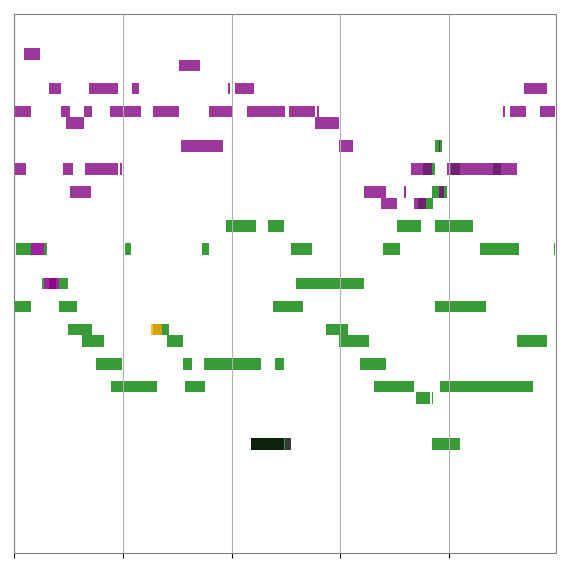}
\caption{$\rightarrow$flu$+$a-g$+$bas}
\label{fig: rollj}
\end{subfigure}~~~
\begin{subfigure}[c]{0.16\linewidth}
\includegraphics[width=\linewidth]{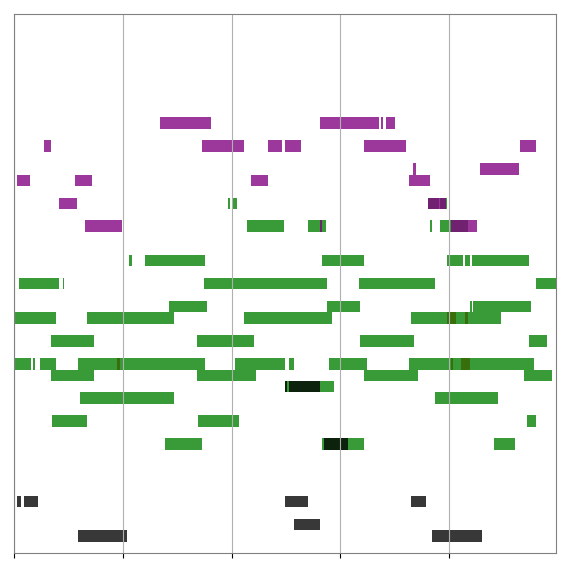}
\caption{\fbox{{flu$+$a-g$+$bas}}}
\label{fig: rollk}
\end{subfigure}
\begin{subfigure}[c]{0.16\linewidth}
\includegraphics[width=\linewidth]{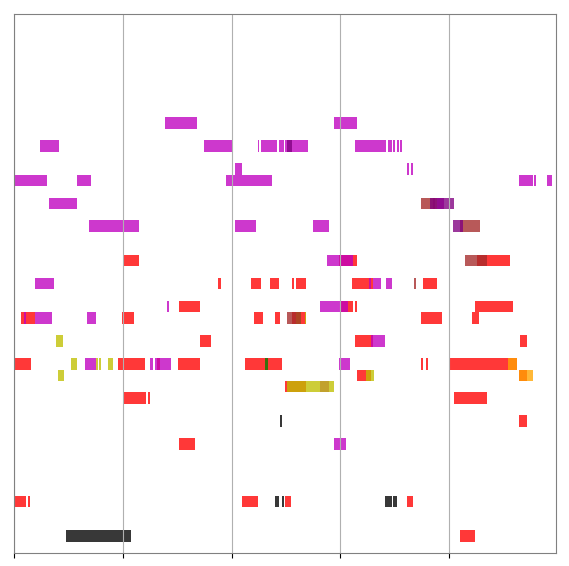}
\caption{$\rightarrow$tru}
\label{fig: rolll}
\end{subfigure}
\caption{Demonstration of timbre crossover (best viewed in color). The source clips are \textbf{(a)}, \textbf{(e)}, \textbf{(g)} and \textbf{(k)}, and the generated ones (i.e., after crossover by UnetAE) are those to the right of them. We see from \textbf{(c)} that UnetAE finds the low-pitched notes for the bass to play; from \textbf{(d)} that it knows to use the violin to play the melody (originally played by the flute) and the acoustic guitar to play the chords (originally played by the piano); and from \textbf{(l)} that it does not always work well---it picks new instruments to play when the 
timbre palettes of the source and target clips do not match. 
[\underline{Purple}: flute (flu), \underline{Red}: piano (pia), \underline{Black}: bass (bas), \underline{Green}: acoustic guitar (a-g), \underline{Light purple}: trumpet (tru), \underline{Yellow}: cello (cel), \underline{Brown}: violin (vio)].} 
\label{fig: 3pianoroll}
\end{figure*}

\subsubsection{Timbre Crossover} 
Secondly, we propose and employ a new evaluation method called \emph{timbre crossover} to evaluate the result of disentanglement. Specifically, we exchange the timbre codes of two existing audio clips and then decode them, to see whether we can exchange their instrumentation without affecting the pitch content in the symbolic domain. 
For example, if clip A (the \emph{source clip}) plays the flute and clip B (the \emph{target clip}) plays the trumpet, we hope that after timbre crossover the new clip A' would use the trumpet to play the original tune.

For objective evaluation, we compare the pitch between the pianorolls of the source clip and the new clip to get the pitch Acc. 
Moreover, we present the activity of different instruments in a clip as an $M$-bin histogram and compute the histogram intersection (HI) \cite{hist_intersection} between the histograms computed from the pianorolls of the target clip and the new clip.
Both pitch accuracy and timbre HI are the higher the better.



Table \ref{tab:pianorolls Acc} tabulates the result of the following crossover scenarios: 
`anything$\rightarrow$piano,' `anything$\rightarrow$guitar,' and \\
`piano$\rightarrow$violin$+$cello.'
We compute the average result for 20 cases for each scenario.
We can see that UnetAE has better pitch accuracy while poorer HI. 
We attribute this to the fact that UnetAE does not have control over the skip connections---some timbre information may still flow through the skip connections. But, the advantage of UnetAE is its timbre code is more pitch-invariant. 
When doing crossover, the pitch content would subject to less changes. 
In contrast, DuoAE achieves higher HI, suggesting that its pitch code is more timbre-invariant. 
But, as its timbre code is not pitch-invariant, when doing crossover, some notes might disappear, due to timbre replacement, 
causing the low pitch accuracy.

Figure \ref{fig: 3pianoroll} demonstrates the result of UnetAE. 
In general, UnetAE works well in changing the timbre without much affecting the pitch.
Please read the caption for details.

Besides, we also adapt and evaluate the models proposed by \cite{liu2018multi} and \cite{hadad2018two} for timbre crossover. For the model proposed by \cite{hadad2018two}, we consider `S' as instrument and `Z' as pitch and use pianorolls as the target output. For \cite{liu2018multi}, we replace `style' with pitch and `class' with instrument. The output is also pianorolls instead of CQT. As Table \ref{tab:pianorolls Acc} shows, they can only achieve 0.558 and 0.295 Pitch Acc for `anything$\rightarrow$piano,' and 0.537 and 0.279 Pitch Acc for `anything$\rightarrow$guitar.' This poor result is expected, since these models were not designed for music disentanglement.

\begin{figure}[t]
\includegraphics[width=.48\linewidth]{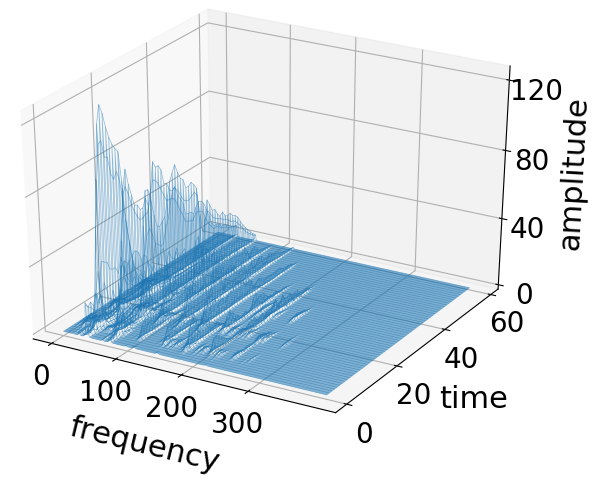}
\includegraphics[width=.48\linewidth]{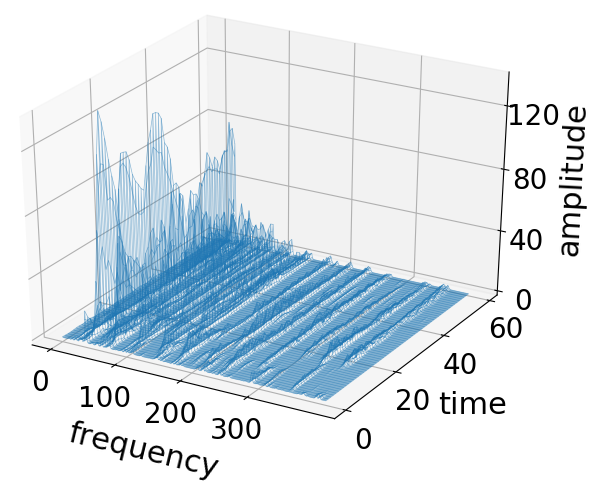}
\caption{The spectrogram of a note from a piano clip (left) and the one after timbre crossover (right) by the UnetAE-based model shown in Figure \ref{fig: editing}. The target timbre is violin.}
\label{fig: stft}
\end{figure}

\subsubsection{Audio-domain Music Editing}

Finally, we demonstrate the result of timbre crossover in the audio domain, using the model depicted in Figure \ref{fig: editing}. 
Figure \ref{fig: stft} shows the spectrograms of a note chunked from the original and generated audio clips, for piano$\rightarrow$violin crossover.
It is known that, compared to piano, violin has longer sustain after the attack, and stronger energy at the harmonics \cite{miyamoto2008harmonic}. We can see such characteristics in the generated spectrogram. However, so far we note that the model may not be sophisticated enough (e.g., perhaps we can additionally use a GAN-based loss) so the audio quality has room for improvement, but it shows that music audio editing is feasible.



\subsection{Ablation Study}

The two models `w/o adv' (i.e., without adversarial training) in Table \ref{tab:pianorolls Acc} can be viewed as ablated versions of the proposed models. We report some more ablation study below.

We replace the pianorolls with CQT as the target output for DuoAE. In our study, we found that predicting pianorolls consistently outperforms predicting CQT in both instrument recognition accuracy (by 9.6\%) and pitch accuracy (by 5.3\%). Using the pianorolls as the target performs much better, for it provides the benefits of ``temporal supervision'' claimed as the first contribution of the paper in our introduction section. We therefore decided to use the pianorolls as the target output for both DuoAE and UnetAE. 

Moreover, we also remove skip connection from the UnetAE. By doing so, the transcription accuracy drops to 0.407. Moreover, since there is no skip connection to help disentangle pitch and timbre, the pitch Acc of timbre crossover would drop from 0.691 to 0.068.

\section{Conclusion and Discussion}

In this paper, we have presented two encoder/decoder models to learn disentangled representation of musical timbre and pitch.
The core contribution is that the input to the models is an audio clip, and as a result it can change the timbre or pitch content of the audio clip by manipulating the learned timbre and pitch representations.
We can evaluate the result of disentanglement by generating new pianorolls or audio clips. 
We also proposed a new evaluation method called ``timbre crossover'' to analyze timbre and pitch disentanglement. Through timbre exchanging, analysis shows that UnetAE has better ability to create new instrumentation without losing pitch information. We also extend UnetAE to audio-domain editing. Result shows that it is feasible to change audio timbre through a deep network structure. 

Although our study shows that UnetAE can perform better than DuoAE in terms of transcription accuracy and timbre crossover, 
a weakness of UnetAE is that it has limited control over the skip connections and as a result part of the timbre information may also flow through skip connection. Besides, UnetAE does not learn a pitch representation, so the model has little control of the pitch information. 

The loss function for adversarial training is another topic for future study. Currently, we predict the zero-metrics $\mathbf{0}_{t}$ and $\mathbf{0}_{p}$ to dispel information from the embeddings. Other loss function, such as the one proposed in \cite{liu2018disentangling}, can also be tested. Besides, it is important to have further evaluation on the timbre embedding. For example, from the timbre crossover we learn that timbre embedding somehow learns the relation between pitch and timbre,
but Figure \ref{fig: rollk} shows that it is not feasible to play the bass and chord with the trumpet. Further experiment can be done to test on variety of music genre and instrument combination. 

The proposed timbre crossover method holds the promise to help human or AI composer decide the instruments to play different parts of a given lead sheet or pianoroll. 
However, a drawback of the current model is that the timbre embeddings have to be picked from another music piece, which is less convenient in real-life usage. A more convenient scenario is that we can interpolate the existing embeddings and directly decide the instruments to use. A possible way to improve the model regarding this is to add one-hot vector for conditional learning. The one-hot vector can then be used for controlling the instrument usage.

The proposed audio editing model, though likely being the first one of its kind, is not sophisticated enough to synthesize realistic audio. The model can be improved by using  WaveNet \cite{van2016wavenet} as the decoder, or by using the multi-band structure proposed by Wang and Yang \cite{wang19aaai}.

\bibliographystyle{IEEE}

\end{document}